\documentstyle{camera}

\begin{document}

%%%%%%%%%%%%%%%%%%%%%%%%%%%%%%%%%%%%%%%%%%%%%%%%%%%%%%%%
% The title, all uppercase; if you want to split it in
% two or more lines, put a \\ macro at each line break
% example: 
%   \title{TITLE: FIRST LINE\\ SECOND LINE}
%

\newcommand{\as}{\alpha_S}

\title{HADRONIC EVENTS AT LEP\thanks{This paper is dedicated to the memory of Frans Verbeure.}}

%%%%%%%%%%%%%%%%%%%%%%%%%%%%%%%%%%%%%%%%%%%%%%%%%%%%%%%%
% The author(s), separated by commas; do not put a
% comma before the last author, use instead the \And
% macro which produces a normal ``and'' in the
% caps/small caps context
%
\author{Alessandro De Angelis}

%%%%%%%%%%%%%%%%%%%%%%%%%%%%%%%%%%%%%%%%%%%%%%%%%%%%%%%%
%
\organization{Dipartimento di Fisica dell'Universit\`a di Udine e INFN Trieste}

\maketitle

%%%%%%%%%%%%%%%%%%%%%%%%%%%%%%%%%%%%%%%%%%%%%%%%%%%%%%%%
% Write the text starting from here and using the usual
% LaTeX commands.
%
\begin{abstract}
The hadronic events collected by LEP between 1989 and 2000 have changed dramatically our understanding of QCD, which has now been established as a part of the Standard Model.
Still in many sectors (nonperturbative in particular) science analysis needs to be completed, and some new studies are just starting: LEP physics is still the best hadronic physics at accelerators, at least in Europe.
\end{abstract}

\section{Introduction}
Studying QCD at LEP has several advantages: the centre-of-mass
energy is high and well defined, jets are collimated, the
environment is clean, statistics is high enough to investigate
even rare topologies, flavour tagging allows the study of jets
originating from light quarks or b- and c-quarks separately, detectors are performant and  measurements can be done over a wide range of energies, reducing systematics.

During the LEP1 phase (1989-1995), where LEP has been operated in the vicinity
of the Z resonance, in total 16 million hadronic events
have been registered by the four LEP collaborations, corresponding to an integrated luminosity around 600 pb$^{-1}$. During the LEP2
phase the centre-of-mass energy has been continuously increased
from 133 GeV to 208 GeV, accumulating an additional integrated luminosity around 2800 pb$^{-1}$. Including data obtained from the analysis
of radiative events from the LEP1 data, the LEP data cover the
energy range from 50 GeV to 208 GeV.

Such an unprecedented amount of hadronic events collected in a clean environment could allow 
precision tests of QCD, the theory of strong interactions. For this purpose, a problem is related to the fact that
a solution of the theory for the observables at hadron level is not yet available.
Further assumptions are thus required: in particular, that quantities that can be reliably computed in perturbation theory correspond to measurable quantities (i.e., that they are robust with respect to hadronization).
Hadronization effects are suppressed by inverse powers of the characteristic energy scale of the process. This assumption, called Local Parton-Hadron Duality (LPHD), implies that a description of a phenomenon in terms of constituents correctly describes the behavior of the hadrons that form the final state (for a review see for example \cite{nason}). 

Between 1989 and today, more than 250 papers on hadronic physics at LEP have been published.
LEP established perturbative QCD as a part of the Standard Model, in particular:
\begin{itemize}
\item measuring the color factors associated to the vertices among the partons, and determining that the measured values are in agreement with QCD (in particular, the self-coupling of the gluon was made evident);
\item measuring the one parameter of the theory: $\as$, which is nowadays conventionally set at the mass of the Z, and determining that its evolution is consistent with QCD. From LEP data only, one has \cite{jones}
\begin{eqnarray*} \as(LEP1) &=& 0.1197
\pm 0.0008(stat,exp)
 \pm 0.0010(had) \pm 0.0048(th)\\
\as(LEP2) &=& 0.1196
\pm 0.0011(stat,exp)
 \pm 0.0007(had) \pm 0.0044(th) 
 \end{eqnarray*}
	(all-energies does not beat LEP II because of high correlation of theory uncertainty).
\end{itemize}

But also  the intermediate regime was developed, transferring the observations at hadron level to the calculations at parton level, via LPHD where possible, or via power corrections, or finally via Monte Carlo models. Among the important results in this "gray area", there was the test of calculations related to multiplicity and momentum spectrum, also for quark-initiated and gluon-initiated jets separately.

Finally, the  fully nonperturbative region was explored, getting insights on particle correlations and final state effects.

And the story is not yet over: around 50 papers are still in the process of finalization by the LEP collaborations.

\section{Two open topics}

In the end, I would like to remind two open topics, who are still very important for the analysis and for which the solution (if found) can be in the mainstream of physics.

\subsection{Interconnection effects in W pairs}

When LEP2 started, the sector of ``soft QCD'' was activated mostly  by the problem of interconnection in WW events.

The problem can be stated as follows. W bosons are mostly produced in pairs at LEP2, and
each W has a probability about 2/3 of decaying into a quark and an antiquark. Since the lifetime of a W, from the Heisenberg principle, is one order of magnitude smaller than the hadronization time, one expects that when both Ws decay hadronically they cannot be treated as independent objects: their decay  diagrams are connected by (soft) gluons (Color Reconnection, CR)
and the hadrons in the final state are mixed together by ``exogamous'' Bose-Einstein Correlations (BEC).

The problem is interesting {\em per se} and in connection with the determination of the W mass:
 the most accurate determination of this quantity can come from the hadronic WW channel, provided we understand interconnection \cite{noiw}.
%\begin{figure}[ht]
%\begin{center}\includegraphics[width=0.4\columnwidth]{ic1.eps}
%\includegraphics[width=0.4\columnwidth]{ic2.eps}
%\end{center}
%\caption{Summary of the LEP results on hadronic interconnection effects in W pairs:
%CR (left) and BEC (right).}
%\end{figure}
 
A surprising prediction was formulated by Andersson and collaborators \cite{bopred}: interconnection in W pairs
could be be very small or zero, since each W hadronizes along a different string and thus they
are two separate objects.
This interpretation attributed a kind of reality of the string concept.

CR and  exogamous BEC have not been established yet: the LEP results are consistent with
the hypothesis of no interconnection at all \cite{intercon}. Until the problem is solved, the data from the fully hadronic decays of W pairs cannot be effectively used for the determination of the W mass.

\subsection{Production of prompt photons}
The decay of an unstable particle into charged particles can be thought as the sudden creation of rapidly moving charges. Such a variation of the electromagnetic field is accompanied by the emission of final state radiation. 
Experiments measure an order of magnitude more radiation than predicted \cite{rad}. 

LEP can establish in a clean environment if this effect can be associated to new fundamental physics \cite{notamia} (parton cascades or maybe partons themselves might be different from what we think).

\section{Conclusions}

Many interesting physics topics are still in the LEP tapes, and the analysis of LEP data can probably offer the best high energy physics in Europe now.  

Students should profit of this circumstance to do real analysis; for this purpose
data should be archived and archiving should be done in the simplest possible way  
(4-vectors are not a shame: they are also used in communities, like NASA, who have a larger experience in making data public).

%%%%%%%%%%%%%%%%%%%%%%%%%%%%%%%%%%%%%%%%%%%%%%%%%%%%%%%%
% End of the paper
%
\end{document}